\documentclass[onecolumn]{svjour3}

\usepackage{graphicx}
\usepackage{bm}
\usepackage{hyperref}

\journalname{Journal of Low Temperature Physics}

\begin{document}

\title{The Fulde-Ferrell-Larkin-Ovchinnikov state in pnictides}
\titlerunning{The Fulde-Ferrell-Larkin-Ovchinnikov state in pnictides}

\author{Andrzej Ptok \and Dawid Crivelli}
\authorrunning{A. Ptok \and D. Crivelli}

\institute{A. Ptok \at 
              Institute of Physics, University of Silesia, 40-007 Katowice, Poland,
              \email{aptok@mmj.pl}
              \and 
              D. Crivelli \at 
              Institute of Physics, University of Silesia, 40-007 Katowice, Poland,
              \email{dcrivelli@us.edu.pl}
}

\date{Received: date / Accepted: date}

\maketitle

%sprawdzic jeszcze raz!
\begin{abstract}
Fe-based superconductors (FeSC) exhibit all the properties of systems that allow the formation of a superconducting phase with oscillating order parameter, called the Fulde--Ferrell--Larkin--Ovchinnikov (FFLO) phase. By the analysis of the Cooper pair susceptibility in two-band FeSC, such systems are shown to support the existence of a FFLO phase, regardless of the exhibited order parameter symmetry.
We also show the state with nonzero Cooper pair momentum, in superconducting FeSC with $\sim \cos(k_{x}) \cdot \cos (k_{y})$ symmetry, to be the ground state of the system in a certain parameter range.
\end{abstract}

\section{Introduction}

At low temperatures the orbital pair breaking effects are smaller in magnitude than the Pauli paramagnetic effect, so that superconductivity survives up to the Pauli limit -- a phase with oscillating order parameter (called the Fulde--Ferrell--Larkin--Ovchinnikov phase or FFLO in short)~\cite{FFLO} can be more stable than a phase with a constant order parameter (the Bardeen--Cooper--Schrieffer phase, or BCS in short). In this case, Cooper pairs may be formed with non-zero total momentum between Zeeman-split parts of the Fermi surface.

Properties of this phase have been usually evaluated in tight-binding models of one-band systems.~\cite{tbmodel} However the latest experimental~\cite{fflo.fesc} and theoretical~\cite{fflo.fesc.th} works suggest we can expect the existence of the FFLO phase in  multi-band Fe-based superconductors (FeSC). It follows from the fact that they possess properties close to heavy fermions systems,~\cite{matsuda.shimahara.07} for which strong experimental evidence suggest the existence of said phase.~\cite{fflo.hf} Both kinds of systems are multi-layered, clean and have a relatively high Maki parameter.

In this paper, making use of the Cooper pair susceptibility and the the minimization of free energy of the system, we discuss the possible appearance of the FFLO phase in pnictides. In Section \ref{sec.theory} we describe the selected model of FeSC, in Section \ref{sec.method} we present our methods. In Section \ref{sec.results} we illustrate and discuss our numerical results. We summarize the results in Section \ref{sec.summary}.

\section{Theoretical model}
\label{sec.theory}

The FeSC system is described using a two-orbital per site model, with hybridization between the $d_{xz}$ and $d_{yz}$ orbitals. We adopt the band structure proposed in Ref.~\cite{raghu.qi.08} and assume that the external magnetic field is parallel to the plane. The Hamiltonian of the system in momentum space takes the following form:
\begin{eqnarray}
\label{eq.ham.mom}
H_{0} &=& \sum_{{\bm k}\sigma} \sum_{\alpha\beta} ( T^{\alpha\beta}_{\bm k} - ( \mu + \sigma h ) \delta_{\alpha\beta} ) c_{\alpha{\bm k}\sigma}^{\dagger} c_{\beta{\bm k}\sigma} \\
\nonumber T^{11}_{\bm k} &=& - 2 \left( t_{1} \cos ( k_{x} ) + t_{2} \cos ( k_{y} ) \right) - 4 t_{3} \cos ( k_{x} ) \cos ( k_{y} ) , \\
\nonumber T^{22}_{\bm k} &=& - 2\left( t_{2} \cos ( k_{x} ) + t_{1} \cos ( k_{y} ) \right) - 4 t_{3} \cos ( k_{x} ) \cos ( k_{y} ) , \\
\nonumber T^{12}_{\bm k} &=& T^{21}_{\bm k} = - 4 t_{4} \sin ( k_{x} ) \sin ( k_{y} ) ,
\end{eqnarray}
where $c_{\alpha{\bm k}\sigma}^{\dagger}$ ($c_{\alpha{\bm k}\sigma}$) is the creation (annihilation) operator of a particle with momentum ${\bm k}$ and spin $\sigma$ in the orbital $\alpha$. $T^{\alpha\beta}_{{\bm k}\sigma}$ is the kinetic energy term of a particle with momentum ${\bm k}$ changing the orbital from $\beta$ to $\alpha$, $\mu$ is the chemical potential and $h$ is the external magnetic field. The hoppings have magnitudes: $(t_{1},t_{2},t_{3},t_{4}) = (-1.0,1.3,-0.85,-0.85)$, in units of $| t_{1} |$. At half-filling, a configurations with two electrons per site requires $\mu = 1.54 | t_{1} |$. Our choice of the parameter set is motivated by the fact that it reproduces the same Fermi surface structure as the local-density approximation calculations of band structure.~\cite{fermisurface}

By diagonalizing the above Hamiltonian, one obtains
\begin{eqnarray}\label{eq.ham.mom}
H'_{0} &=& \sum_{\varepsilon{\bm k}\sigma} E_{\varepsilon{\bm k}\sigma} d_{\varepsilon{\bm k}\sigma}^{\dagger} d_{\varepsilon{\bm k}\sigma}
\end{eqnarray}
with eigenvalues $E_{\varepsilon{\bm k}\sigma} = E_{\varepsilon{\bm k}} - ( \mu + \sigma h )$, where:
\begin{equation}
E_{\pm,{\bm k}} = \frac{ T_{\bm k}^{11} + T_{\bm k}^{22} }{2} \pm \sqrt{ \left( \frac{ T_{\bm k}^{11} - T_{\bm k}^{22} }{2} \right)^{2} + \left( T_{\bm k}^{12} \right)^{2} } ,
\end{equation}
$d_{\varepsilon{\bm k}\sigma}^{\dagger}$ is a new fermion quasi-particle operator in the band $\varepsilon = \pm$. In this case we have two Fermi surfaces (Fig. \ref{fig.sgnOP}.a) -- giving an electron-like band ($\varepsilon=+$) and hole-like band ($\varepsilon=-$).

\section{Methods}
\label{sec.method}

%%%%%%%%%%%%%%%%%%%%%%%%%%%%%%%%%%%%%%%%%%%%%%%%%%%%%%%%%%
\begin{figure}
\begin{center}
\includegraphics{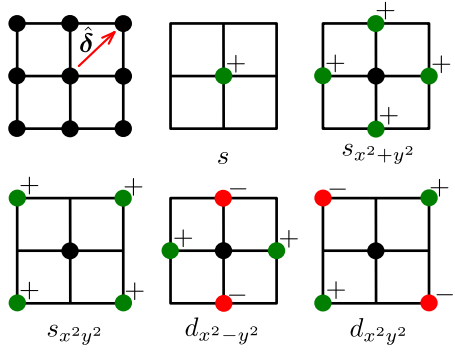}
\end{center}
\caption{(Color online) The vector ${\bm \delta}$ defines the pairing between sites $i$ and $i+{\bm \delta}$ for different symmetries of the order parameter. Colors and symbols correspond to the sign of the order parameter for a given direction in real space. For {\it s-wave} symmetry the pairing is between two electrons on the same site of the lattice, while for other symmetries it is between two other sites (nearest neighbors or next nearest neighbors). In contrast to $d$ type symmetries, $s$ type symmetries do not change sign depending on the direction.}
\label{fig.sym}
\end{figure}
%%%%%%%%%%%%%%%%%%%%%%%%%%%%%%%%%%%%%%%%%%%%%%%%%%%%%%%%%%

We introduce a superconducting pairing between the long-lived quasi-particles in bands $\varepsilon = \pm$.~\cite{linder.sudbo.09} To determine the possibility of formation of the FFLO phase, we turn our attention to the static Cooper pairs susceptibility:
\begin{eqnarray}\label{eq.podpar.def}
\chi_{\varepsilon}^{\Delta} ( {\bm q} ) &\equiv & \lim_{\omega \rightarrow 0} \frac{-1}{N} \sum_{{\bm i}{\bm j}} \exp ( i {\bm q} \cdot ( {\bm i} - {\bm j} ) ) \langle \langle \widehat{\Delta}_{\varepsilon {\bm i}} | \widehat{\Delta}_{\varepsilon {\bm j}}^{\dagger} \rangle \rangle^{r} ,
\end{eqnarray}
where $\langle \langle \ldots \rangle \rangle^{r}$ is the retarded Green's function and $\widehat{\Delta}_{\varepsilon {\bm i}} = \sum_{\bm j} \vartheta ( {\bm j} - {\bm i} ) d_{\varepsilon {\bm i} \uparrow} d_{\varepsilon {\bm j} \downarrow}$ is the OP in band $\varepsilon$. The operator $d_{\varepsilon{\bm i} \sigma}$ in real space corresponds to the operator $d_{\varepsilon{\bm k} \sigma}$ in momentum space. The Factor $\vartheta ( {\bm j} - {\bm i} )$ defines the OP symmetries (Fig. \ref{fig.sym}) -- for example for $d_{x^{2}-y^{2}}${\it -wave} pairing, $\vartheta ( {\bm \delta} )$ is equal to $1$ ($-1$) for ${\bm \delta} = \pm \hat{x}$ ($\pm \hat{y}$) and zero otherwise.
%%%%%%%%%%%%%%%%%%%%%%%% < tu byl rysunek
In momentum space:
\begin{eqnarray}
\label{eq.podpar.mom} \chi_{\varepsilon}^{\Delta} ( {\bm q} ) = \lim_{\omega \rightarrow 0} \frac{-1}{N} \sum_{{\bm k}{\bm l}} \eta ( -{\bm k}-{\bm q} ) \eta ( {\bm l} ) {\bm G}_{\varepsilon} ( {\bm k},{\bm l},{\bm q},\omega ) ,
\end{eqnarray}
\begin{eqnarray}
\nonumber  {\bm G}_{\varepsilon} ( {\bm k},{\bm l},{\bm q},\omega ) = \langle \langle d_{\varepsilon {\bm k} \uparrow} d_{\varepsilon, -{\bm k}-{\bm q} \downarrow} | d_{\varepsilon , -{\bm l}-{\bm q} \downarrow}^{\dagger} d_{\varepsilon {\bm l} \uparrow}^{\dagger} \rangle \rangle^{r} = \delta_{{\bm k}{\bm l}} \frac{ f ( - E_{\varepsilon {\bm k} \uparrow} ) - f ( E_{\varepsilon, -{\bm k}-{\bm q} \downarrow} ) }{\omega - E_{\varepsilon {\bm k} \uparrow} - E_{\varepsilon, -{\bm k}-{\bm q} \downarrow}} , \\
\end{eqnarray}
where $\eta ( {\bm k} )$ is the structure factor:
\begin{eqnarray}
\eta ( {\bm k} ) = \left \lbrace \begin{array}{cc}
1 & $for s-wave$ \\ 
2 \left( \cos( k_{x} ) + \cos ( k_{y} ) \right) & $for $s_{x^{2}+y^{2}}$-wave$ , \\ 
4 \cos( k_{x} ) \cos ( k_{y} ) & $for $s_{x^{2}y^{2}}(s_\pm)$-wave$ , \\ 
2 \left( \cos( k_{x} ) - \cos ( k_{y} ) \right) & $for $d_{x^{2}-y^{2}}$-wave$ , \\ 
4 \sin( k_{x} ) \sin ( k_{y} ) & $for $d_{x^{2}y^{2}}$-wave$ , 
\end{array} \right.
\end{eqnarray}
corresponding to the type of symmetry of the OP.

We investigate the tendency to form the FFLO phase in the system, using the static Cooper pairs susceptibility $\chi_{\varepsilon}^{\Delta} ( {\bm q} )$. 
In magnetic fields of the order of the Pauli limit, when the critical FFLO field ($h_{c}^{FFLO}$) is bigger than the corresponding BCS field ($h_{c}^{BCS}$), the FFLO phase is favored. In such case, the divergence of this function for some ${\bm q} \neq 0$ may imply a second-order transition to the FFLO state of corresponding symmetry from the normal phase.~\cite{mierzejewski.ptok.09}
The location of the maximum of the response function $\chi_{\varepsilon}^{\Delta} ( {\bm q} )$ matches the preferred momentum of the Cooper pairs in the system described by the Hamiltonian (\ref{eq.ham.mom}) in magnetic field $h$. This method allows to establish the propensity to form the superconducting phase (with non-zero momentum of the Cooper pairs) without specifying the mechanisms responsible for the ordered phases with given symmetry. Additionally we obtain the change in the pair susceptibilities $\delta \chi_{\varepsilon}^{\Delta} ( {\bm q} ) = \chi_{\varepsilon}^{\Delta} ( {\bm q} ) - \bar{\chi}_{\varepsilon}^{\Delta} ( {\bm q} )$ due to the external magnetic field ($\chi_{\varepsilon}^{\Delta} ( {\bm q} )$ with the field, $\bar{\chi}_{\varepsilon}^{\Delta}  ( {\bm q} )$ without respectively).

It should be noted that the divergence of the Cooper-pair susceptibility is neither a sufficient condition nor evidence for the transition to the FFLO state. In order for this to happen the system energy $\Omega ( {\bm q} )$ should attain its minimum at a nonzero Cooper pair momentum ${\bm q}$ in a magnetic field $h > h_{c}^{BCS}$, equivalent to the condition $h_{c}^{FFLO} > h_{c}^{BCS}$. To check this, we effectively describe superconductivity in the FFLO phase by the Hamiltonian:
\begin{equation}
H_{SC} = \sum_{\varepsilon {\bm k}} \left( \Delta_{\varepsilon{\bm k}} d_{\varepsilon{\bm k}\uparrow} d_{\varepsilon,-{\bm k}+{\bm q}_{\varepsilon} \downarrow} + H.c. \right) ,
\end{equation}
where $\Delta_{\varepsilon{\bm k}} = \Delta_{\varepsilon} \eta( {\bm k} )$ is the amplitude of the OP for Cooper pairs with total momentum ${\bm q}_{\varepsilon}$ (in band $\varepsilon$ with symmetry described by $\eta ({\bm k})$).
As we see, in the operator basis $d_{\varepsilon{\bm k}\sigma}$ the total Hamiltonian $H = H'_{0} + H_{SC}$ formally describes a system with two independent bands.
Using the Bogoliubov transformation we can find the eigenvalues of $H$:
\begin{eqnarray}
\lambda_{\varepsilon{\bm k}}^{\pm} &=& \frac{E_{\varepsilon{\bm k}\uparrow} - E_{\varepsilon,-{\bm k}+{\bm q}\downarrow}}{2} \pm \sqrt{ \left( \frac{E_{\varepsilon{\bm k}\uparrow} + E_{\varepsilon,-{\bm k}+{\bm q}\downarrow}}{2} \right)^{2} + | \Delta_{\varepsilon{\bm k}} |^{2} } .
\end{eqnarray}
The free energy is given by:
\begin{eqnarray}
\Omega = - k T \sum_{\alpha \in \pm} \sum_{\varepsilon{\bm k}} \ln \left( 1 + \exp ( - \beta \lambda_{\varepsilon{\bm k}}^{\alpha} ) \right) + \sum_{\varepsilon {\bm k}} \left( E_{\varepsilon {\bm k} \downarrow} - \frac{ 2 | \Delta_{\varepsilon} |^{2} }{ V_{\varepsilon} } \right) ,
\end{eqnarray}
where $V_{\varepsilon}$ is the interaction intensity in band $\varepsilon$. The global ground state for fixed $h$ and $T$ is found by minimizing the free energy w.r.t. the OPs and ${\bm q}$.

\section{Numerical results and discussion}
\label{sec.results}

Numerical calculations were carried out for a square lattice $N_{x} \times N_{y} = 600 \times 600$ with periodic boundary conditions, for $kT = 10^{-5} | t_{1} |$. As a first step the static Cooper pairs susceptibility $\chi_{\varepsilon}^{\Delta} ( {\bm q} )$ was calculated in magnetic field $h = 0.25 | t_{1} |$. Then the free energy $\Omega ( {\bm q} )$ of the superconducting system was evaluated for magnetic fields near the Pauli limit $h_{P} \simeq 0.25 | t_{1} |$.

\begin{figure*}
\includegraphics{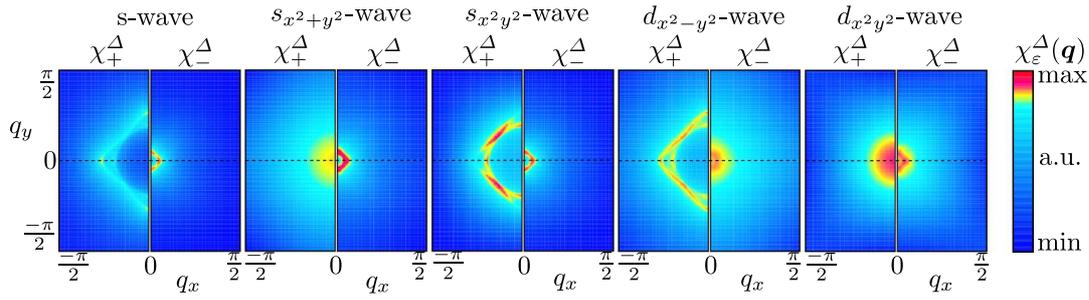}
\caption{(Color online) The static Cooper pairs susceptibility $\chi_{\varepsilon}^{\Delta}  ( {\bm q} )$ in the presence of the external magnetic field $h = 0.25 | t_{1} |$ and $kT = 10^{-5} | t_{1} |$ for different symmetries.}
\label{fig.pod1}
\end{figure*}

\paragraph{The static Cooper pairs susceptibility.} Assuming different symmetries $\eta ( {\bm k } )$ of the superconducting OP in bands $\varepsilon = \pm$, we characterized the Cooper pair susceptibility -- Fig. \ref{fig.pod1}. For every OP symmetry, in the band $\varepsilon=-$ the static Cooper pairs susceptibility $\chi_{-}^{\Delta} ( {\bm q} )$ takes its maximum for ${\bm q} \neq 0$. Conversely in the band $\varepsilon=+$, with $s_{x^{2}+y^{2}}$ and $d_{x^{2}y^{2}}$ symmetry of the OP, there is a strong tendency to form a BCS phase (maximum $\chi_{+}^{\Delta} ( {\bm q} )$ for ${\bm q} = 0$). When $h_{c}^{FFLO} > h_{c}^{BCS}$ this can be a sign of the existence of the FFLO phase in the band $\varepsilon=-$, while the band $\varepsilon=+$ is in the normal state. Numerical data for both {\it d-wave} type symmetries in band $\varepsilon = -$ is less clear cut, as the maximum $\chi ( {\bm q} )$ is only slightly greater than  $\chi ( {\bm 0} )$.

%%%%%%%%%%%%%%%%%%%%%%%%%%%%%%%%%%%%%%%%%%%%%%%%%%%%%%%%%%
%%%%%%%%%%%%%%%%%%%%%%%%%%%%%%%%%%%%%%%%%%%%%%%%%%%%%%%%%%

%%%%%%%%%%%%%%%%%%%%%%%%%%%%%%%%%%%%%%%%%%%%%%%%%%%%%%%%%%
\begin{figure*}
\includegraphics{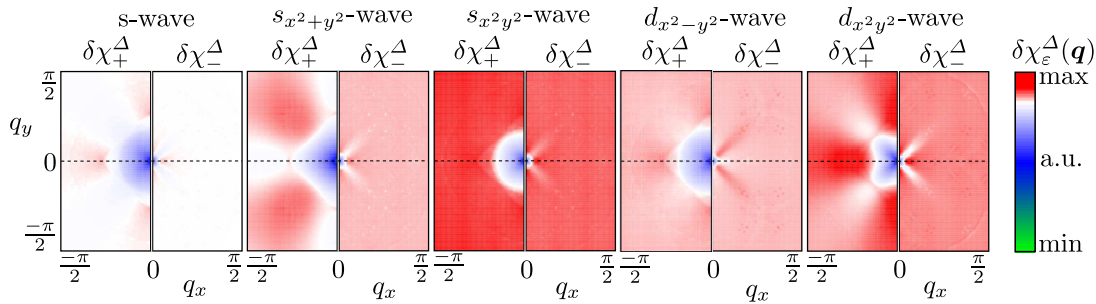}
\caption{(Color online) Change in the static Cooper pairs susceptibility $\delta \chi_{\varepsilon}^{\Delta}  ( {\bm q} )$ (for data presented in Fig. \ref{fig.pod1})
%, $h = 0.25 | t_{1} |$  and $kT = 10^{-5} | t_{1} |$) for different symmetries.
}
\label{fig.pod2}
\end{figure*}
%%%%%%%%%%%%%%%%%%%%%%%%%%%%%%%%%%%%%%%%%%%%%%%%%%%%%%%%%%

There is a clear preference in case of $\varepsilon=-$ for much smaller momenta than in band $\varepsilon=+$, due to the relative width of the bands. Cooper pair momenta depend on the split in the Fermi surface, caused by the external magnetic field, which is larger for the broader band $\varepsilon=+$. Additionally the presence of a magnetic field causes a dampening in each case of the response function near zero momentum (Fig. \ref{fig.pod2} -- in blue). Nonetheless larger momenta are unaffected and increasing (in red).

The behaviour of the response function $\chi_{\varepsilon}^{\Delta} ( {\bm q} )$ shows that multi-band systems have the characteristics typical of one-band systems -- Cooper pairs in the FFLO phase possessing momentum along the principal directions of the system are preferred,~\cite{fflo.onedirection} -- for example in directions $[\pm 1,0]$ and $[\pm 1,1]$ for {\it s-wave} and $d_{x^{2}-y^{2}}$-{\it wave} symmetry respectively in band $\varepsilon=+$  (Fig. \ref{fig.pod1}).

\begin{figure*}
\begin{center}
\includegraphics{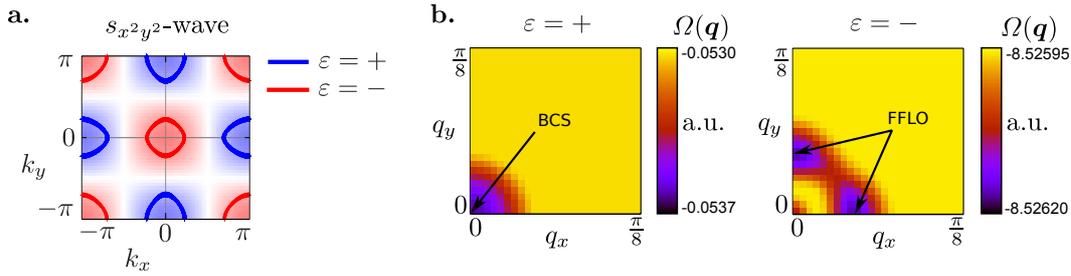}
\end{center}
\caption{
(Color online) Detailed study of the minimal two-band model describing iron-base superconductors with 
$s_{x^{2}y^{2}}(s_{\pm})${\it -wave} symmetry proposed by Ref.~\cite{raghu.qi.08}.
(Panel a) Fermi surfaces (solid line) for $\mu = 1.54 | t_{1} |$. The background color describes the sign of the OP (red for $\eta ( {\bm k} ) > 0$ and blue for $\eta ( {\bm k} ) < 0$). 
(Panel b) The free energy $\Omega ( {\bm q} )$ in the two bands $\varepsilon = \pm$, 
for different values of the Cooper pair momentum ${\bm q}$, 
showing the location of the minima and indicating the existence of different phases. 
Results for $h = 0.25 | t_1 |$ and temperature $kT = 10^{-5} | t_{1} |$.
}
\label{fig.sgnOP}
\end{figure*}

\paragraph{Minimization of free energy.} Theoretical results indicate the presence of $s_{x^{2}y^{2}} \sim \cos ( k_{x} ) \cdot \cos ( k_{y} )$ (also called $s_{\pm}$) pairing symmetry in FeSC .~\cite{op.symmetry} In this case the OP exhibits a sign reversal between the hole pockets and electron pockets (Fig. \ref{fig.sgnOP}.a). Taking this into account, in this paragraph only consider $s_{x^{2}y^{2}}$ symmetry. $V_{\varepsilon}$ was taken such that the Pauli limit was of the order $h_{P} \simeq 0.25 | t_{1} |$ ($V_{+} = -0.74 | t_{1} |$ and $V_{-} = - 1.56 | t_{1} |$).

%%%%%%%%%%%%%%%%%%%%%%%%%%%%%%%%%%%%%%%%%%%%%%%%%%%%%%%%%%
%%%%%%%%%%%%%%%%%%%%%%%%%%%%%%%%%%%%%%%%%%%%%%%%%%%%%%%%%%

The study of the free energy $\Omega ( {\bm q} )$ for the BCS state (${\bm q} = 0 $) w.r.t. magnetic fields $h \simeq h_{P}$, showed that phase transitions in both bands are first-order for all symmetries, except for $s_{x^{2} + y^{2}}$ and $d_{x^{2}y^{2}}$ which are second-order. Only the minimization of $\Omega ( {\bm q} )$ w.r.t. ${\bm q}$, allows to check whether the system exhibits a FFLO phase. Varying ${\bm q} \in FBZ$ in case of $s_{x^{2}y^{2}}$ pairing showed that the band $\varepsilon = +$ undergoes a transition from BCS to the normal state and the band $\varepsilon = -$ from BCS to FFLO state (Fig. \ref{fig.sgnOP}.b). Further increasing the magnetic field, the FFLO phase persists in $\varepsilon = -$. It should be pointed out that in this band exist four equivalent Cooper pair momenta $( \pm q , 0 )$ and $( 0, \pm q )$, in agreement with the static Cooper pairs susceptibility results, and also with previous works.~\cite{fflo.onedirection} Moreover, it is reasonable
  to expect that the phase with an OP given by the superposition of plane waves with said momenta would be energetically favored by the system.~\cite{moreq}

\section{Summary}
\label{sec.summary}

FeSC exhibit many characteristic features of systems in which we can expect the existence of the FFLO phase. Using a minimal two-band model for FeSC, we conducted a numerical study of FFLO phase in multi-band systems.
The static Cooper pair susceptibility suggests that we can expect the system to prefer the state with nonzero Cooper pair momenta (the FFLO phase) regardless of the OP symmetry, when $h_{c}^{FFLO} > h_{c}^{BCS}$.
Moreover, the ground state of the system with $s_{x^{2}y^{2}} \sim \cos(k_{x}) \cdot \cos (k_{y})$ symmetry OP, can be the state with nonzero Cooper pair momentum for magnetic fields near the Pauli limit.

\begin{acknowledgements}
D.C. acknowledges a scholarship from the TWING project, co-funded by the European Social Fund.
\end{acknowledgements}

\end{document}